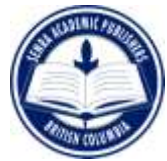

# ON DISCOVERY OF EXTRA FOUR NEW DISPERSIVE SH-WAVES IN MAGNETOELECTROELASTIC PLATES

A.A Zakharenko
International Institute of Zakharenko Waves (IIZWs)
660014, ul. Chaikovskogo, 20-304, Krasnoyarsk, Russia

## ABSTRACT

The study of this theoretical work definitely adds extra four new dispersive shear-horizontal waves propagating in the transversely isotropic piezoelectromagnetic (PEM) plates of class 6 *mm*. In this study, the following mechanical, electrical, and magnetic boundary conditions at both the upper and lower free surfaces of the PEM plate are exploited: the mechanically free surface, continuity of both the electrical and magnetic inductions, and continuity of both the electrical and magnetic potentials. The obtained dispersion relations were also graphically studied and compared with previous results. Some interesting peculiarities were also discussed. It is well-known that the plate waves are frequently used for nondestructive testing and evaluation of thin films, further miniaturization of different technical devices based on PEM smart materials, and constitution of new technical devices such as filters, sensors, delay lines, switches, lab-on-a-chip, etc.

## INTRODUCTION

Development of modern (free of the charge) electronics called spintronics requires intensive implementation of new phenomena. It is predicted that spintronics can completely substitute the convectional electronics indoor only in several decades in the best case. It is expected that some found new phenomena can even accelerate the spintronics realization. The time period between the discovery of some competitive phenomena and their realization in some smart technical devices must be also shortened. This theoretical report relates to the first problem of creation of new phenomena in order that the research community can be familiar with them and therefore, can incorporate them in modern electronics by the way of creation of novel technical devices based on some suitable new phenomena. It is thought that a class of the magnetoelectric smart materials such as piezoelectromagnetics (magnetoelectroelastics) is actually apt for spintronics and can even provide more possibilities for revealing new phenomena. This report has a purpose to introduce new knowledge coupled with the discovery of new wave phenomena. It is obvious that this knowledge can be obtained only by the way of development of theoretical knowledge. The theory developed below can be compared with the previously found phenomena (Zakharenko, 2015a; Zakharenko, 2015b; Zakharenko, 2015c; Zakharenko, 2013a) and actually enriches theoretical knowledge. This knowledge was not reviewed in the single existing review by Zakharenko (2013b) on the wave phenomena in piezoelectromagnetics and represents the additional discovery to those obtained in book by Zakharenko (2012a).

It was recently found in theoretical paper (Zakharenko, 2015b) that the magnetoelectric effect characterized by the electromagnetic constant $α$ can demonstrate a dramatic influence on the existence of the shear-horizontal surface acoustic waves (SH-SAWs) propagating in the transversely isotropic magnetoelectroelastics. However, the surface waves require the utilization of bulk piezoelectromagnetics (magnetoelectroelastics) that can be used in constitution of smart technical devices with a high level of integration. Further device miniaturization requires some incorporation of smart material thin films or plates. The plate SH-waves can be used in filters, sensors, actuators, delay lines, lab-on-a-chip, etc. Indeed, piezoelectromagnetic smart materials can provide co-influence between the electrical and magnetic subsystems through the mechanical subsystem. This can explain their

---

Corresponding author e-mail: aazaaz@inbox.ru

growing popularity in various applications. However, it is frequently met that some strongly piezoelectromagnetic materials can have some bulk, surficial, and interfacial defects, i.e. they can be brittle. Therefore, complex non-destructive testing and evaluation of such materials by ultrasonic waves can be irreplaceable. Also, these materials are promising for active control of sound and vibration, ultrasonic medical imaging, etc. They can categorically have desirable frequency response, high resolution, and generate large forces. This is highly-called for modern technologies and aerospace industry to constitute transducers such as actuators and sensors. Also, wireless sensorics (Durdag, 2009) is the other intrigues subject of application of smart materials because SH-waves can possess the highest sensitivity. It is already well-known that piezoelectromagnetics instead of conventional piezoelectrics are preferable for noncontact generation and detection of different SH-waves. The noncontact method (Hirao and Ogi, 2003; Ribichini *et al*., 2010; Thompson, 1990) based on the electromagnetic acoustic transducer (EMAT) is the right one to study the magnetoelectric SH-wave phenomena that can be found in bulk magnetoelectroelastics and two-dimensional samples. This research report has no purpose to mention all the promising possibilities for piezoelectromagnetic (composite) materials. Some properties of such smart magnetoelectric materials and their applications can be found in selected review papers (Fiebig, 2005; Özgür *et al*., 2009; Pullar, 2012; Srinivasan, 2010). Also, it is necessary to mention that piezoelectromagnetic composites possessing stronger magnetoelectric effect are preferable for commercial applications in comparison with piezoelectromagnetic monocrystals. However, recent discovery of ferroelectric monocrystals (Kimura, 2012) possessing commercially reasonable magnetoelectric effect states that composites are not single solution. The study of this report does not require that composites or monocrystal must have a strong magnetoelectric effect. It will be demonstrated below that a material with a fittingly weak magnetoelectric effect can be preferable. Indeed, this weak effect can significantly slow down the new SH-waves and even act on the wave existence (Zakharenko, 2015b).

It is also indispensable to mention the famous SH-SAWs existing in the transversely isotropic piezoelectrics or piezomagnetics. They are known as the surface Bleustein-Gulyaev (BG) waves (Bleustein, 1968; Gulyaev, 1969) theoretically discovered to the end of the 1960s. The propagation of this type of the nondispersive SH-SAWs is supported by the piezoelectric or piezomagnetic effect in piezoelectrics or piezomagnetics, respectively. It is obvious that the surface BG-waves do not relate to the magnetoelectric effect. Piezoelectromagnetics can actually possess simultaneously all of these three mentioned effects such as the piezoelectric, piezomagnetic, and magnetoelectric effects. As a result, some rivalry among the effects leads to the possible existence of more than ten new SH-SAWs (Zakharenko, 2015b; Zakharenko, 2015c; Zakharenko, 2013a; Zakharenko, 2013b; Zakharenko, 2010a; Zakharenko, 2011; Zakharenko, 2012b; Melkumyan, 2007). However, only several of them can be associated with the key influence of the magnetoelectric effect. Indeed, as soon as the electromagnetic constant $\alpha$ is equal to zero, some piezoelectromagnetic new SH-SAWs cannot exist. This fact occurs for the fifth new SH-SAW discovered by Zakharenko (2010a) and further studied in (Zakharenko, 2011), eighth and tenth new SH-SAWs (Zakharenko, 2015b; Zakharenko, 2015c; Zakharenko, 2013a). For $\alpha = 0$, the fifth new SH-SAW cannot propagate. However, it was demonstrated in (Zakharenko, 2015b) that the dependence of the eighth and tenth new SH-SAWs is more dramatic because small values of the electromagnetic constant $\alpha$ can radically slow down their propagation speeds. Moreover, these waves cannot exist for nonzero values of the $\alpha$.

This theoretical report discovers several new dispersive SH-waves propagating in the transversely isotropic magnetoelectroelastic plates. Some of the plate new SH-waves relate to the tenth new SH-SAW in the limit case when the plate thickness approaches an infinity (the case of bulk sample.) A set of technical applications still requires knowledge of the wave properties of bulk and thin-film magnetoelectroelastic (composite) samples. Thus, the following section acquaints the reader with the theoretical background leading to the existence of new solutions.

**Theoretical foundation**

Proper choice of the thermodynamic functions and thermodynamic variables allows constitution of suitable consistent equations (Zakharenko, 2012a) for a piezoelectromagnetic solid. For this case there can be chosen the following apt three thermodynamic functions: stress $\sigma_{ij}$, electrical induction $D_i$, and magnetic induction $B_i$, where the indices $i$ and $j$ run from 1 to 3. For the case of small perturbations, three independent thermodynamic variables are: the strain $\eta_{ij} = 0.5(\partial U_i/\partial x_j + \partial U_j/\partial x_i)$, electrical field $E_i = -\partial \varphi/\partial x_i$, and magnetic field $H_i = -\partial \psi/\partial x_i$, where $U_i$ and $x_i$ are the components of the mechanical displacements and real space, respectively. Also, $\varphi$ and $\psi$ are the electrical and magnetic potentials. These thermodynamic functions and variables thermodynamically define all the material constants of the piezoelectromagnetics. As a result, it is possible to constitute the coupled equations of motion in the quasi-static approximation because the speed of the electromagnetic waves is approximately five orders larger than that of the acoustic waves. For this purpose, the governing equations for mechanical equilibrium, electrostatic equilibrium, and magnetostatic equilibrium must be written. Finally, three coupled equations of motion in the differential form are as follows: $\partial \sigma_{ij}/\partial x_j = \rho \partial^2 U_i/\partial t^2$, $\partial D_i/\partial x_j = 0$, and $\partial B_i/\partial x_j = 0$, where $\rho$ and $t$ are the material mass density and time, respectively. These equations can be used for theoretical investigations of the in-plane and anti-plane polarized wave in any possible propagation direction of a piezoelectromagnetics of any known symmetry. However, it is useful to exploit the plane wave solutions to rewrite the equations of motion in the tensor form (Zakharenko, 2012a). Let's treat the concrete case of the SH-wave propagation in the transversely isotropic piezoelectromagnetic plate.

To study SH-wave propagation, the transversely isotropic (6 $mm$) materials are popular for theoretical and experimental investigations because many suitable cuts and propagation directions can be found in the solids. The suitable propagation directions must satisfy the condition of perpendicularity to the sixfold axis of crystal symmetry (Gulyaev, 1998). Regarding to the SH-waves, they must be polarized along the sixfold symmetry axis. These conditions are true for bulk and thin-film samples. Let's consider the acoustic wave propagation managed along the $x_1$-axis of the rectangular coordinate system. For the SH-waves, the polarization must be directed along both the $x_2$-axis and the sixfold symmetry axis. The $x_3$-axis coincides with the normal to the cut surface. The coordinate beginning is situated at the middle of the plate thickness. So, the upper and lower surfaces of the piezoelectromagnetic plate can be reached at $x_3 = + d$ and $x_3 = - d$, respectively.

The considered propagation direction allows the existence of pure waves (Auld, 1990; Dieulesaint and Royer, 1980; Lardat *et al.*, 1971). This means that the propagation of pure SH-wave is coupled with both the electrical and magnetic potentials and there also propagates the purely mechanical Rayleigh wave with the in-plane polarization. This is possible because in this case the coupled equations of motion written in the convenient tensor form (modified Green-Christoffel equation) are separated into two independent parts (Zakharenko, 2012a). This report has an interest in investigation of the propagation of pure SH-wave in the piezoelectromagnetic plate. Therefore, only the following part of the equations of motion can be taken into account:

$$\begin{pmatrix} C[p - (V_{ph}/V_{t4})^2] & ep & hp \\ ep & -\varepsilon p & -\alpha p \\ hp & -\alpha p & -\mu p \end{pmatrix} \begin{pmatrix} U^0 \\ \varphi^0 \\ \psi^0 \end{pmatrix} = \begin{pmatrix} 0 \\ 0 \\ 0 \end{pmatrix} \quad (1)$$

where $p = 1 + m_3^2$ and $V_{t4} = \text{sqrt}(C/\rho)$ is the speed of purely mechanical bulk SH-wave; $\rho$ is the material mass density and $m_3$ is the one of the directional cosines: $m_1 = 1$, $m_2 = 0$, and $m_3 = m_3$. Also, the phase velocity $V_{ph}$ must be found, $V_{ph} = \omega/k$, where $\omega$ and $k$ are the angular frequency and the wavenumber in the propagation direction.

In this set of three homogeneous equations written in matrix form (1), the following material parameters (Zakharenko, 2010a; Zakharenko, 2012a; Zakharenko, 2013a) contribute: the elastic stiffness constant $C$, piezoelectric constant $e$, piezomagnetic coefficient $h$, dielectric permittivity coefficient $\varepsilon$, magnetic permeability coefficient $\mu$, and electromagnetic constant $\alpha$. Also, equation (1) contains the eigenvector components such as $U^0$, $\varphi^0$, and $\psi^0$. They must be found for each known value of the eigenvalue $m_3$. Expanding the determinant associated with the square matrix of coefficients in equation (1), a sextic polynomial in a single indeterminate $m_3$ can be written. This obtained secular homogeneous equation can reveal the following six polynomial roots:

$$m_3^{(1,3)} = -m_3^{(2,4)} = \mp j, \quad m_3^{(5,6)} = \mp j\sqrt{1 - (V_{ph}/V_{tem})^2} \quad (2)$$

It is clearly seen that the fifth and sixth polynomial roots depend on the shear-horizontal bulk acoustic wave (SH-BAW denoted by $V_{tem}$) that is coupled with both the electrical ($\varphi$) and magnetic ($\psi$) potentials and defined by

$$V_{tem} = \sqrt{C/\rho}\left(1 + K_{em}^2\right)^{1/2} \quad (3)$$

where

$$K_{em}^2 = \frac{\mu e^2 + \varepsilon h^2 - 2\alpha eh}{C(\varepsilon\mu - \alpha^2)} = \frac{e(e\mu - h\alpha) - h(e\alpha - h\varepsilon)}{C(\varepsilon\mu - \alpha^2)} \quad (4)$$

It is natural that SH-BAW velocity (3) is a function of the coefficient of the magnetoelectromechanical coupling (CMEMC) defined by expression (4). It is demonstrated that the CMEMC contains three coupling mechanisms such as $(e\alpha - h\varepsilon)$, $(e\mu - h\alpha)$, and $(\varepsilon\mu - \alpha^2)$. They were recently discussed in short report by Zakharenko (2013c).

With equations' set (1), each found eigenvalue $m_3$ can provide the corresponding eigenvector. The suitable forms of the eigenvectors are discussed in paper by Zakharenko (2014a). Thence, the determined explicit forms of the components of the first four eigenvectors (Zakharenko, 2012a; Zakharenko, 2010a) read:

$$\begin{pmatrix} U^{0(1)} \\ \varphi^{0(1)} \\ \psi^{0(1)} \end{pmatrix} = \begin{pmatrix} U^{0(2)} \\ \varphi^{0(2)} \\ \psi^{0(2)} \end{pmatrix} = \begin{pmatrix} U^{0(3)} \\ \varphi^{0(3)} \\ \psi^{0(3)} \end{pmatrix} = \begin{pmatrix} U^{0(4)} \\ \varphi^{0(4)} \\ \psi^{0(4)} \end{pmatrix} = \begin{pmatrix} 0 \\ \mu \\ -\alpha \end{pmatrix} \quad (5)$$

The fifth and sixth sets of the eigenvector components are more complicated. They can be expressed as follows:

$$\begin{pmatrix} U^{0(5)} \\ \varphi^{0(5)} \\ \psi^{0(5)} \end{pmatrix} = \begin{pmatrix} U^{0(6)} \\ \varphi^{0(6)} \\ \psi^{0(6)} \end{pmatrix} = \begin{pmatrix} (e\mu - h\alpha)/CK_{em}^2 \\ \mu - h^2/CK_{em}^2 \\ -\alpha + eh/CK_{em}^2 \end{pmatrix}$$

$$= \frac{1}{K_{em}^2} \begin{pmatrix} (e\mu - h\alpha)/C \\ \mu(K_{em}^2 - K_m^2) \\ -\alpha(K_{em}^2 - K_\alpha^2) \end{pmatrix} = \frac{e\mu - h\alpha}{CK_{em}^2(\varepsilon\mu - \alpha^2)} \begin{pmatrix} \varepsilon\mu - \alpha^2 \\ e\mu - h\alpha \\ -(e\alpha - h\varepsilon) \end{pmatrix} \quad (6)$$

where

$$K_{em}^2 - K_m^2 = \frac{(e\mu - h\alpha)^2}{C\mu(\varepsilon\mu - \alpha^2)} \quad (7)$$

$$K_{em}^2 - K_\alpha^2 = \frac{(e\alpha - h\varepsilon)(e\mu - h\alpha)}{C\alpha(\varepsilon\mu - \alpha^2)} \quad (8)$$

$$K_m^2 = \frac{h^2}{C\mu} \quad (9)$$

$$K_\alpha^2 = \frac{eh}{C\alpha} = \frac{\alpha eh}{C\alpha^2} \quad (10)$$

It is obvious that these three coupling mechanisms of the CMEMC mentioned after expression (4) are present in eigenvectors (6) and the other expressions such as equalities (7) and (8). They will be met in the further analysis. They also represent important coupling mechanisms that must compete with each other. In expression (9), the nondimensional parameter $K_m^2$ is called the coefficient of the magnetomechanical coupling (CMMC). This coefficient is also used to characterize a pure piezomagnetics. The other nondimensional parameter denoted by $K_\alpha^2$ (10) contains the electromagnetic constant $\alpha$. The reader can also check that eigenvectors (5) and (6) are coupled via the second coupling mechanism such as $(e\mu - h\alpha)$. This can be illuminated by the following equalities:

$$e\varphi^{0(1)} + h\psi^{0(1)} = e\varphi^{0(5)} + h\psi^{0(5)} = e\mu - h\alpha \quad (11)$$

The determined eigenvalues and eigenvectors allow composition of the complete parameters. These complete parameters can be naturally written in the plane wave forms. They are called the complete mechanical displacement $U^\Sigma$ directed along the $x_2$-axis, complete electrical potential $\varphi^\Sigma$, and complete magnetic potential $\psi^\Sigma$. Using the index $I = 2, 4, 5$, it si possible to write down that $U_2^\Sigma = U^\Sigma$, $U_4^\Sigma = \varphi^\Sigma$, and $U_5^\Sigma = \psi^\Sigma$. Thus, these three complete parameters can be compactly written in the following plane wave form:

$$U_I^\Sigma = \sum_{q=1,2,3,4,5,6} F^{(q)} U_I^{0(q)} \exp[jk(m_1 x_1 + m_3^{(q)} x_3 - V_{ph} t)] \quad (12)$$

where $j = (-1)^{1/2}$ is the imaginary unity.

In expression (12), the exponent can be introduced via the hyperbolic cosine and hyperbolic sine: $\exp(\pm\Theta) = \cosh(\Theta) \pm \sinh(\Theta)$. Thence, it is possible to rewrite the complete parameters for the case of $V_{ph} < V_{tem}$. Therefore, the parameters $U^\Sigma$, $\varphi^\Sigma$, and $\psi^\Sigma$ can be inscribed in the following convenient form:

$$U_I^\Sigma = \{F_{01} U_I^{0(1)} \cosh(kx_3) + F_{02} U_I^{0(1)} \sinh(kx_3) \quad (13)$$
$$+ F_{03} U_I^{0(5)} \cosh\left(kx_3 \sqrt{1 - (V_{ph}/V_{tem})^2}\right)$$
$$+ F_{04} U_I^{0(5)} \sinh\left(kx_3 \sqrt{1 - (V_{ph}/V_{tem})^2}\right)\} \times \exp[jk(x_1 - V_{ph} t)]$$

where $F_{01} = F^{(1)} + F^{(2)} + F^{(3)} + F^{(4)}$, $F_{02} = F^{(1)} + F^{(2)} - F^{(3)} - F^{(4)}$, $F_{03} = F^{(5)} + F^{(6)}$, and $F_{04} = F^{(5)} - F^{(6)}$.

It is necessary to state that expressions (12) and (13) are only applicable within the plate thickness, $-d \leq x_3 \leq +d$. In these expressions, the corresponding weight factors can be determined from the applied mechanical, electrical, and magnetic boundary conditions that can be applied at both the sides of the piezoelectromagnetic plate. Let's assume that the upper ($x_3 = +d$) and lower ($x_3 = -d$) surfaces of the transversely isotropic PEM plate are in a contact with a vacuum. Therefore, the material parameters of the latter continuum must be also treated. For a vacuum, the dielectric permittivity and magnetic permeability constants are $\varepsilon_0 = 0.08854187817 \times 10^{-10}$ [F/m] and $\mu_0 = 1.25663706144 \times 10^{-6}$ [N/A$^2$], respectively. Using the subscript "$f$" for the free space (vacuum) it is potential to employ the well known Laplace equations $\Delta\varphi_f = 0$ and $\Delta\psi_f = 0$ rewritten as follows: $(k_1^2 + k_3^2)\varphi_{f0} = 0$ and $(k_1^2 + k_3^2)\psi_{f0} = 0$. The electrical and magnetic potentials above the upper surface ($x_3 = +d$) of the PEM can be written in the following forms: $\varphi_{f0} = F^{(E0)} \exp(-k_1 x_3) \exp[j(k_1 x_1 - \omega t)]$ and $\psi_{f0} = F^{(M0)} \exp(-k_1 x_3) \exp[j(k_1 x_1 - \omega t)]$, $x_3 \geq +d$. Below the lower surface ($x_3 = -d$) they are: $\varphi_{f0} = F^{(E0)} \exp(k_1 x_3) \exp[j(k_1 x_1 - \omega t)]$ and $\psi_{f0} = F^{(M0)} \exp(k_1 x_3) \exp[j(k_1 x_1 - \omega t)]$, $x_3 \leq -d$. Also, both the potentials must exponentially decay in a vacuum when $x_3 > +d$ and $x_3 < -d$.

Let's use homogeneous boundary conditions at the upper ($x_3 = +d$) and lower ($x_3 = -d$) surfaces of the PEM plate. This means that they are the same at either surface. The mechanical traction-free boundary condition at the free surfaces relates to the normal component of the stress tensor: $\sigma_{32} = 0$. The electrical and magnetic boundary conditions are the continuity of both the electrical and magnetic potentials ($\varphi = \varphi^f$ and $\psi = \psi^f$) and the continuity of the normal component of both the electrical and magnetic displacements ($D_3 = D^f$ and $B_3 = B^f$) where the superscript "$f$" relates to the free space (vacuum.) All the possible boundary conditions are perfectly described in Al'shits *et al.* (1992).

Next, it is natural to perfume some transformations to exclude the vacuum weight factors $F_E$ and $F_M$ in order to deal only with six homogeneous equations in six unknowns $F^{(1)}$, $F^{(2)}$, $F^{(3)}$, $F^{(4)}$, $F^{(5)}$, and $F^{(6)}$. Using equations (1) and (13), these six homogeneous equations corresponding to the boundary conditions can be rewritten as those in only four unknowns $F_{01}$, $F_{02}$, $F_{03}$, and $F_{04}$. With $b = \sqrt{1 - (V_{ph}/V_{tem})^2}$, these six homogeneous equations (three ones for the upper surface at $x_3 = +d$ and the rest three for the lower surface at $x_3 = -d$) are written as follows:

$$(CU^{0(1)} + e\varphi^{0(1)} + h\psi^{0(1)})[F_{01}\sinh(kd) + F_{02}\cosh(kd)]$$
$$+ b(CU^{0(5)} + e\varphi^{0(5)} + h\psi^{0(5)})[F_{03}\sinh(bkd) + F_{04}\cosh(bkd)] = 0$$
(14)

$$-(eU^{0(1)} - \varepsilon\varphi^{0(1)} - \alpha\psi^{0(1)})\{F_{01}\sinh(kd) + F_{02}\cosh(kd)\}$$
$$-b(eU^{0(5)} - \varepsilon\varphi^{0(5)} - \alpha\psi^{0(5)})\{F_{03}\sinh(bkd) + F_{04}\cosh(bkd)\}$$
$$+\varepsilon_0\{F_{01}\varphi^{0(1)}\cosh(kd) + F_{02}\varphi^{0(1)}\sinh(kd)$$
$$+ F_{03}\varphi^{0(5)}\cosh(bkd) + F_{04}\varphi^{0(5)}\sinh(bkd)\} = 0$$
(15)

$$-(hU^{0(1)} - \alpha\varphi^{0(1)} - \mu\psi^{0(1)})\{F_{01}\sinh(kd) + F_{02}\cosh(kd)\}$$
$$-b(hU^{0(5)} - \alpha\varphi^{0(5)} - \mu\psi^{0(5)})\{F_3\sinh(bkd) + F_{04}\cosh(bkd)\}$$
$$+\mu_0\{F_{01}\psi^{0(1)}\cosh(kd) + F_{02}\psi^{0(1)}\sinh(kd)$$
$$+ F_{03}\psi^{0(5)}\cosh(bkd) + F_{04}\psi^{0(5)}\sinh(bkd)\} = 0$$
(16)

$$(CU^{0(1)} + e\varphi^{0(1)} + h\psi^{0(1)})[F_{01}\sinh(kd) - F_{02}\cosh(kd)]$$
$$+ b(CU^{0(5)} + e\varphi^{0(5)} + h\psi^{0(5)})[F_{03}\sinh(bkd) - F_{04}\cosh(bkd)] = 0$$
(17)

$$-(eU^{0(1)} - \varepsilon\varphi^{0(1)} - \alpha\psi^{0(1)})\{F_{01}\sinh(kd) - F_{02}\cosh(kd)\}$$
$$-b(eU^{0(5)} - \varepsilon\varphi^{0(5)} - \alpha\psi^{0(5)})\{F_{03}\sinh(bkd) - F_{04}\cosh(bkd)\}$$
$$+\varepsilon_0\{F_{01}\varphi^{0(1)}\cosh(kd) - F_{02}\varphi^{0(1)}\sinh(kd)$$
$$+ F_{03}\varphi^{0(5)}\cosh(bkd) - F_{04}\varphi^{0(5)}\sinh(bkd)\} = 0$$
(18)

$$-(hU^{0(1)} - \alpha\varphi^{0(1)} - \mu\psi^{0(1)})\{F_{01}\sinh(kd) - F_{02}\cosh(kd)\}$$
$$-b(hU^{0(5)} - \alpha\varphi^{0(5)} - \mu\psi^{0(5)})\{F_{03}\sinh(bkd) - F_{04}\cosh(bkd)\}$$
$$+\mu_0\{F_{01}\psi^{0(1)}\cosh(kd) - F_{02}\psi^{0(1)}\sinh(kd)$$
$$+ F_{03}\psi^{0(5)}\cosh(bkd) - F_{04}\psi^{0(5)}\sinh(bkd)\} = 0$$
(19)

To further transform these equations, eigenvector components (5) and (6) together with the following equalities are useful:

$$CU^{0(1)} + e\varphi^{0(1)} + h\psi^{0(1)} = e\mu - h\alpha \tag{20}$$

$$CU^{0(5)} + e\varphi^{0(5)} + h\psi^{0(5)} = (e\mu - h\alpha)\frac{1 + K_{em}^2}{K_{em}^2} \tag{21}$$

$$eU^{0(1)} - \varepsilon\varphi^{0(1)} - \alpha\psi^{0(1)} = -\varepsilon\mu + \alpha^2 \tag{22}$$

$$eU^{0(5)} - \varepsilon\varphi^{0(5)} - \alpha\psi^{0(5)}$$
$$= \frac{\mu e^2 - \alpha eh}{CK_{em}^2} + \frac{\varepsilon h^2}{CK_{em}^2} - \varepsilon\mu - \frac{\alpha eh}{CK_{em}^2} + \alpha^2 = 0 \tag{23}$$

$$hU^{0(1)} - \alpha\varphi^{0(1)} - \mu\psi^{0(1)} = -\alpha\mu + \mu\alpha = 0 \tag{24}$$

$$hU^{0(5)} - \alpha\varphi^{0(5)} - \mu\psi^{0(5)}$$
$$= \frac{e\mu h - \alpha h^2}{CK_{em}^2} + \frac{\alpha h^2}{CK_{em}^2} - \alpha\mu - \frac{e\mu h}{CK_{em}^2} + \alpha\mu = 0 \tag{25}$$

After the transformations, these six homogeneous equations take the following forms:

$$(e\mu - h\alpha)\{F_{01}\sinh(kd) + F_{02}\cosh(kd)$$
$$+ b\frac{1 + K_{em}^2}{K_{em}^2}[F_{03}\sinh(bkd) + F_{04}\cosh(bkd)]\} = 0 \tag{26}$$

$$(\varepsilon\mu - \alpha^2)\{F_{01}\sinh(kd) + F_{02}\cosh(kd)\}$$
$$+\varepsilon_0\mu\{F_{01}\cosh(kd) + F_{02}\sinh(kd)$$
$$+ \frac{K_{em}^2 - K_m^2}{K_{em}^2}[F_{03}\cosh(bkd) + F_{04}\sinh(bkd)]\} = 0 \tag{27}$$

$$\mu_0\alpha\{F_{01}\cosh(kd) + F_{02}\sinh(kd)$$
$$+ \frac{K_{em}^2 - K_\alpha^2}{K_{em}^2}[F_{03}\cosh(bkd) + F_{04}\sinh(bkd)]\} = 0 \tag{28}$$

$$(e\mu - h\alpha)\{F_{01}\sinh(kd) - F_{02}\cosh(kd)$$
$$+ b\frac{1 + K_{em}^2}{K_{em}^2}[F_{03}\sinh(bkd) - F_{04}\cosh(bkd)]\} = 0 \tag{29}$$

$$(\varepsilon\mu-\alpha^2)\{F_{01}\sinh(kd)-F_{02}\cosh(kd)\}$$
$$+\varepsilon_0\mu\{F_{01}\cosh(kd)-F_{02}\sinh(kd)$$
$$+\frac{K_{em}^2-K_m^2}{K_{em}^2}[F_{03}\cosh(bkd)-F_{04}\sinh(bkd)]\}=0 \quad (30)$$

$$\mu_0\alpha\{F_{01}\cosh(kd)-F_{02}\sinh(kd)$$
$$+\frac{K_{em}^2-K_\alpha^2}{K_{em}^2}[F_{03}\cosh(bkd)-F_{04}\sinh(bkd)]\}=0 \quad (31)$$

It is a standard procedure for the plate wave study that the obtained six homogeneous equations in four unknowns $F_{01}$, $F_{02}$, $F_{03}$, and $F_{04}$ written above in the final forms must be separated into two independent sets of homogeneous equations. They are three equations in unknowns $F_{01}$ and $F_{03}$ and the other three equations in unknowns $F_{02}$ and $F_{04}$. This is possible because there are three pairs of equations that can be transformed: equations (26) and (29), (27) and (30), and (28) and (31). So, the first three equations read:

$$(e\mu-h\alpha)\left\{F_{01}\sinh(kd)+b\frac{1+K_{em}^2}{K_{em}^2}F_{03}\sinh(bkd)\right\}=0 \quad (32)$$

$$(\varepsilon\mu-\alpha^2)F_{01}\sinh(kd)$$
$$+\varepsilon_0\mu\left\{F_{01}\cosh(kd)+\frac{K_{em}^2-K_m^2}{K_{em}^2}F_{03}\cosh(bkd)\right\}=0 \quad (33)$$

$$\mu_0\alpha\left\{F_{01}\cosh(kd)+\frac{K_{em}^2-K_\alpha^2}{K_{em}^2}F_{03}\cosh(bkd)\right\}=0 \quad (34)$$

The second independent set of three homogeneous equations then read:

$$(e\mu-h\alpha)\left\{F_{02}\cosh(kd)+b\frac{1+K_{em}^2}{K_{em}^2}F_{04}\cosh(bkd)\right\}=0 \quad (35)$$

$$(\varepsilon\mu-\alpha^2)F_{02}\cosh(kd)$$
$$+\varepsilon_0\mu\left\{F_{02}\sinh(kd)+\frac{K_{em}^2-K_m^2}{K_{em}^2}F_{04}\sinh(bkd)\right\}=0 \quad (36)$$

$$\mu_0\alpha\left\{F_{02}\sinh(kd)+\frac{K_{em}^2-K_\alpha^2}{K_{em}^2}F_{04}\sinh(bkd)\right\}=0 \quad (37)$$

It is obvious that these two independent sets of three homogeneous equations in two unknowns must provide some solutions. For this purpose, equations (32), (33), and (34) must be consistent from each other, namely one of three equations must represent a sum of the rest two. The same must occur for the second independent set of three equations from (35) to (37). Therefore, the following two sections provide new solutions of new dispersive SH-waves propagating in the piezoelectromagnetic plates.

**The first pair of new dispersive SH-waves**

This case relates to the third coupling mechanism of the CMEMC such as $(\varepsilon\mu-\alpha^2)$. It is natural to multiply equations (32) and (34) by the factors of $(\varepsilon\mu-\alpha^2)/(e\mu-h\alpha)$ and $\varepsilon_0\mu/\mu_0\alpha$, respectively. Therefore, the first set of the equations can be readily transformed into the following one:

$$(\varepsilon\mu-\alpha^2)\left\{F_{01}\sinh(kd)+b\frac{1+K_{em}^2}{K_{em}^2}F_{03}\sinh(bkd)\right\}=0 \quad (38)$$

$$(\varepsilon\mu-\alpha^2)F_{01}\sinh(kd)$$
$$+\varepsilon_0\mu\left\{F_{01}\cosh(kd)+\frac{K_{em}^2-K_m^2}{K_{em}^2}F_{03}\cosh(bkd)\right\}=0 \quad (39)$$

$$\varepsilon_0\mu\left\{F_{01}\cosh(kd)+\frac{K_{em}^2-K_\alpha^2}{K_{em}^2}F_{03}\cosh(bkd)\right\}=0 \quad (40)$$

It is clearly seen that these three equations are now actually not independent from each other. Indeed, they are consistent because equation (39) represents a sum of equations (38) and (40). Consequently, these two equations must be successively subtracted from main equation (39). This procedure results in the following dispersion relation, namely the dependence of the velocity $V_{new37}$ of the thirty seventh new SH-wave on the normalized plate half-thickness $kd$:

$$\sqrt{1-(V_{new37}/V_{tem})^2}\tanh\left(kd\sqrt{1-(V_{new37}/V_{tem})^2}\right)$$
$$-\frac{\varepsilon_0\mu}{\varepsilon\mu-\alpha^2}\frac{K_\alpha^2-K_m^2}{1+K_{em}^2}=0 \quad (41)$$

The same simple transformations can be also applied to the second set of the equations. Therefore, the second three equations also become consistent and read as follows:

$$(\varepsilon\mu-\alpha^2)\left\{F_{02}\cosh(kd)+b\frac{1+K_{em}^2}{K_{em}^2}F_{04}\cosh(bkd)\right\}=0 \quad (42)$$

$$(\varepsilon\mu-\alpha^2)F_{02}\cosh(kd)$$
$$+\varepsilon_0\mu\left\{F_{02}\sinh(kd)+\frac{K_{em}^2-K_m^2}{K_{em}^2}F_{04}\sinh(bkd)\right\}=0 \quad (43)$$

$$\varepsilon_0\mu\left\{F_{02}\sinh(kd)+\frac{K_{em}^2-K_\alpha^2}{K_{em}^2}F_{04}\sinh(bkd)\right\}=0 \quad (44)$$

A successive subtraction of equations (42) and (44) from main equation (43) definitely leads to a new dispersion relation. The dependence of the velocity $V_{new38}$ of the thirty eighth new SH-wave on the normalized parameter $kd$ can be introduced as follows:

$$\sqrt{1-(V_{new38}/V_{tem})^2} - \frac{\varepsilon_0\mu}{\varepsilon\mu-\alpha^2}\frac{K_\alpha^2-K_m^2}{1+K_{em}^2}\tanh\left(kd\sqrt{1-(V_{new38}/V_{tem})^2}\right)=0 \quad (45)$$

One can check that in the limit case of $kd \to \infty$, i.e. the hyperbolic tangent approaches the unity, both the obtained dispersion relations reduce to the relatively simple formula for calculation of the nondispersive tenth SH-SAW velocity recently discovered in theoretical work Zakharenko (2015c). This velocity is defined by

$$V_{new10}=V_{tem}\left[1-\left(\frac{\varepsilon_0\mu}{\varepsilon\mu-\alpha^2}\frac{K_m^2-K_\alpha^2}{1+K_{em}^2}\right)^2\right]^{1/2}$$

$$=V_{tem}\left[1-\left(-\frac{h\varepsilon_0}{e\alpha}\frac{e(e\mu-h\alpha)}{C\varepsilon\mu-C\alpha^2+\mu e^2+\varepsilon h^2-2\alpha eh}\right)^2\right]^{1/2} \quad (46)$$

**The second pair of new dispersive SH-waves**

It is obvious that this case relates to the second coupling mechanism of the CMEMC such as $(e\mu-h\alpha)$. This means that equation (32) stays unchanged. Besides, rest equations (33) and (34) must be properly transformed. These transformations are also simple to get the case of consistent equations. Therefore, the first three equations can be written down in the following final forms:

$$(e\mu-h\alpha)\left\{F_{01}\sinh(kd)+b\frac{1+K_{em}^2}{K_{em}^2}F_{03}\sinh(bkd)\right\}=0 \quad (47)$$

$$e\mu F_{01}\sinh(kd)+\frac{e\varepsilon_0\mu^2}{\varepsilon\mu+\varepsilon_0\mu-\alpha^2}$$
$$\times\left\{F_{01}[\cosh(kd)-\sinh(kd)]+\frac{K_{em}^2-K_m^2}{K_{em}^2}F_{03}\cosh(bkd)\right\}=0 \quad (48)$$

$$-h\alpha F_{01}\sinh(kd)$$
$$-h\alpha\left\{F_{01}[\cosh(kd)-\sinh(kd)]+\frac{K_{em}^2-K_\alpha^2}{K_{em}^2}F_{03}\cosh(bkd)\right\}=0 \quad (49)$$

It is possible to use equation (47) to release the final dispersion relation for this case. Indeed, equation (47) can be subtracted from a sum of equations (48) and (49). This results in a complicated dispersion relation. Thus, the propagation velocity $V_{new39}$ of the thirty ninth new SH-wave in the piezoelectromagnetic plate can be calculated from the following dispersion relation:

$$0=\sqrt{1-\left(\frac{V_{new39}}{V_{tem}}\right)^2}\tanh\left(kd\sqrt{1-\left(\frac{V_{new39}}{V_{tem}}\right)^2}\right)\tanh(kd)+\frac{h\alpha}{e\mu-h\alpha}$$
$$\times\left\{\sqrt{1-\left(\frac{V_{new39}}{V_{tem}}\right)^2}\tanh\left(kd\sqrt{1-\left(\frac{V_{new39}}{V_{tem}}\right)^2}\right)[\tanh(kd)-1]+\frac{K_{em}^2-K_\alpha^2}{1+K_{em}^2}\tanh(kd)\right\}$$
$$-\frac{e\mu}{e\mu-h\alpha}\frac{\varepsilon_0\mu}{\varepsilon\mu+\varepsilon_0\mu-\alpha^2}$$
$$\times\left\{\sqrt{1-\left(\frac{V_{new39}}{V_{tem}}\right)^2}\tanh\left(kd\sqrt{1-\left(\frac{V_{new39}}{V_{tem}}\right)^2}\right)[\tanh(kd)-1]+\frac{K_{em}^2-K_m^2}{1+K_{em}^2}\tanh(kd)\right\} \quad (50)$$

To obtain the dispersion relation written above, the following relation between the amplitude coefficients $F_{01}$ and $F_{03}$ borrowed from equation (47) was used:

$$F_{01}=-b\frac{1+K_{em}^2}{K_{em}^2}F_{03}\frac{\sinh(bkd)}{\sinh(kd)} \quad (51)$$

It is now necessary to transform in the same manner the second set of three homogeneous equations from (35) to (37). As a consequence, these three equations take the following forms:

$$(e\mu-h\alpha)\left\{F_{02}\cosh(kd)+b\frac{1+K_{em}^2}{K_{em}^2}F_{04}\cosh(bkd)\right\}=0 \quad (52)$$

$$e\mu F_{02}\cosh(kd)+\frac{e\varepsilon_0\mu^2}{\varepsilon\mu+\varepsilon_0\mu-\alpha^2}$$
$$\times\left\{F_{02}[\sinh(kd)-\cosh(kd)]+\frac{K_{em}^2-K_m^2}{K_{em}^2}F_{04}\sinh(bkd)\right\}=0 \quad (53)$$

$$-h\alpha F_{02}\cosh(kd)$$
$$-h\alpha\left\{F_{02}[\sinh(kd)-\cosh(kd)]+\frac{K_{em}^2-K_\alpha^2}{K_{em}^2}F_{04}\sinh(bkd)\right\}=0 \quad (54)$$

These three equations written above are also consistent. Their consistency leads to an extra dispersion relation. Here, the same mathematical procedure plays a role that was applied in order to obtain the previous dispersion relation. Accordingly, the velocity $V_{new40}$ of the fortieth new SH-wave can be evaluated from the following equation:

$$\sqrt{1-\left(\frac{V_{new40}}{V_{tem}}\right)^2}+\frac{h\alpha}{e\mu-h\alpha}$$

$$\times\left\{\sqrt{1-\left(\frac{V_{new40}}{V_{tem}}\right)^2}\left[1-\tanh(kd)\right]+\frac{K_{em}^2-K_\alpha^2}{1+K_{em}^2}\tanh\left(kd\sqrt{1-\left(\frac{V_{new40}}{V_{tem}}\right)^2}\right)\right\}$$

$$-\frac{e\mu}{e\mu-h\alpha}\frac{\varepsilon_0\mu}{\varepsilon\mu+\varepsilon_0\mu-\alpha^2}$$

$$\times\left\{\sqrt{1-\left(\frac{V_{new40}}{V_{tem}}\right)^2}\left[1-\tanh(kd)\right]+\frac{K_{em}^2-K_m^2}{1+K_{em}^2}\tanh\left(kd\sqrt{1-\left(\frac{V_{new40}}{V_{tem}}\right)^2}\right)\right\}=0$$

(55)

In this dispersion relation, dependence (52) between the $F_{02}$ and $F_{04}$ was used. This can be defined by

$$F_{02}=-b\frac{1+K_{em}^2}{K_{em}^2}F_{04}\frac{\cosh(bkd)}{\cosh(kd)}\qquad(56)$$

It is possible to analytically investigate the limit case of $kd\to\infty$. This case significantly simplifies both dispersion relations (50) and (55). The resulting equation serves for evaluation of the eleventh new SH-SAW velocity recently discovered by Zakharenko (2015c). So, it is practical to write down this formula below:

$$V_{new11}=V_{tem}\left[1-\left(-\frac{h\alpha}{e\mu-h\alpha}\frac{K_{em}^2-K_\alpha^2}{1+K_{em}^2}+\frac{e\mu}{e\mu-h\alpha}\frac{\varepsilon_0\mu}{(\varepsilon+\varepsilon_0)\mu-\alpha^2}\frac{K_{em}^2-K_m^2}{1+K_{em}^2}\right)^2\right]^{1/2}$$

$$=V_{tem}\left[1-\left(\frac{\varepsilon_0\mu e(e\mu-h\alpha)-(\varepsilon\mu+\varepsilon_0\mu-\alpha^2)h(e\alpha-h\varepsilon)}{(\varepsilon\mu+\varepsilon_0\mu-\alpha^2)(C\varepsilon\mu-C\alpha^2+\mu e^2+\varepsilon h^2-2\alpha eh)}\right)^2\right]^{1/2}$$

(57)

It is decisive to state that this theoretical study has discovered four new dispersion relations given by formulae (41), (45), (50), and (55). Some of them look like complicated expressions. Therefore, it is required to graphically illustrate the velocity dependencies on the value of $kd$ by the means of the use of real piezoelectromagnetic composite materials.

**Comparative graphical study**

The chosen piezoelectromagnetic composites for the further study are $BaTiO_3$–$CoFe_2O_4$ and PZT-5H–Terfenol-D. The material parameters for the first composite are: $\rho=5730$ [kg/m$^3$], $C=4.4\times10^{10}$ [N/m$^2$], $e=5.8$ [C/m$^2$], $h=275.0$ [T], $\varepsilon=56.4\times10^{-10}$ [F/m], $\mu=81.0\times10^{-6}$ [N/A$^2$]. Those for the second are: $\rho=8500$ [kg/m$^3$], $C=1.45\times10^{10}$ [N/m$^2$], $e=8.5$ [C/m$^2$], $h=83.8$ [T], $\varepsilon=75.0\times10^{-10}$ [F/m], $\mu=2.61\times10^{-6}$ [N/A$^2$]. These material parameters were also used in several studies (Zakharenko, 2013d; Zakharenko, 2013e; Zakharenko, 2014b). The utilization of the same material parameters can be useful because it allows comparison of different results. This graphical study is based on computation of dispersion relations (41), (45), (50), and (55) obtained in this theoretical work. The calculated dependencies of the propagation velocities on the nondimensional parameter $kd$ are graphically shown in Figures 1 and 2, where $k$ is the wavenumber in the propagation direction and $d$ is the plate half-thickness.

Figures 1a and 1b show dispersion relations (41) and (45), namely the fundamental modes of the dispersive SH-waves propagating in the PEM plates for the studied piezoelectromagnetic composite materials PZT-5H–Terfenol-D and $BaTiO_3$–$CoFe_2O_4$, respectively. The dispersion relations are calculated for various values of the nondimensional parameter $\alpha^2/\varepsilon\mu$. It is clearly seen in figure 1 that for both the studied magnetoelectroelastic composites there exist a "silence" zone for the normalized velocity $V_{new37}/V_{tem}$ (thick lines) because the velocity $V_{new37}$ starts with its zero value at some nonzero value of the parameter $kd$. One can find in figures 1a and 1b that the larger silence zone occurs for smaller value of the parameter $\alpha^2/\varepsilon\mu$ because the velocity $V_{new37}$ has to start with a larger value of the $kd$. Concerning the normalized velocity $V_{new38}/V_{tem}$ also shown in the figures by the thinner lines, it can start at a small value of the $kd$ when $V_{new38}>V_{tem}$ can occur. However, this study has an interest in the case of $V_{new38}<V_{tem}$ and therefore, only this case is shown in the figures. At large values of the $kd$, the values of both the dispersive SH-wave velocities $V_{new37}$ and $V_{new38}$ approach the value of nondispersive SH-SAW velocity $V_{new10}$ (46) that is clearly seen in the figures. There is also one peculiarity for these SH-waves propagating with the velocities $V_{new37}$, $V_{new38}$, and $V_{new10}$: they can exist when the value of the parameter $\alpha^2/\varepsilon\mu$ is larger than some threshold value denoted by $(\alpha^2/\varepsilon\mu)_{th}$. For the studied composites, these threshold values are small: $(\alpha^2/\varepsilon\mu)_{th}\sim4.8\times10^{-8}$ for PZT-5H–Terfenol-D and $(\alpha^2/\varepsilon\mu)_{th}\sim5.0\times10^{-9}$ for $BaTiO_3$–$CoFe_2O_4$.

This peculiarity mentioned above can be used for constitution of some technical devices and also exist in the other study developed by Zakharenko (2015a). It is possible to discuss that some piezoelectromagnetic switches can be manufactured exploiting this peculiarity of the SH-wave existence: the SH-wave propagation for some suitable value of $\alpha^2/\varepsilon\mu>(\alpha^2/\varepsilon\mu)_{th}$ can represent the ON-regime and $\alpha^2/\varepsilon\mu<(\alpha^2/\varepsilon\mu)_{th}$ without any SH-wave propagation can originate the OFF-regime. These regimes can be also appropriate for the computer logics or data storage devices because the ON- and OFF-regimes can correspond to "1" and "0" states, or vice versa, respectively. An external magnetic field as an example can be responsible to cause and control the aforementioned regimes. Also, it is blatant that some values of $\alpha^2/\varepsilon\mu$ being slightly larger than the value of $(\alpha^2/\varepsilon\mu)_{th}$ can represent an interest for the reader regarding very slow PEM-SH-wave propagation with the speeds $V_{new37}$ and $V_{new38}$. It is hoped that very slow speeds of the acoustic SH-waves coupled with both the electrical and magnetic potentials can be experimentally studied.

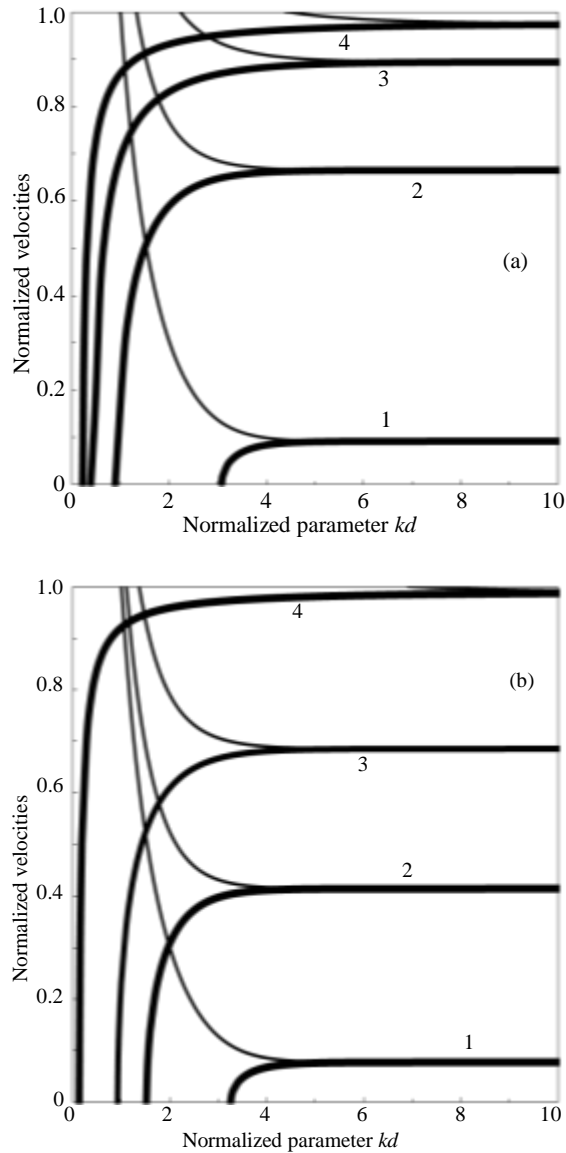

Fig. 1. The normalized velocities $V_{new37}/V_{tem}$ (thick lines, formula (41)) and $V_{new38}/V_{tem}$ (thinner lines, formula (45)) of the fundamental modes of the dispersive SH-waves propagating in the PEM plates versus the normalized value of the half-thickness $kd$: (a) PZT-5H–Terfenol-D, where (1) $\alpha^2/\varepsilon\mu = 5.0625 \times 10^{-8}$, (2) $3.0 \times 10^{-8}$, (3) $2.5 \times 10^{-7}$, (4) $1.0 \times 10^{-6}$; (b) BaTiO$_3$–CoFe$_2$O$_4$, where (1) $\alpha^2/\varepsilon\mu = 5.329 \times 10^{-9}$, (2) $6.4 \times 10^{-9}$, (3) $1.0 \times 10^{-8}$, (4) $2.5 \times 10^{-7}$.

It is expected that properly performed experiments will allow researchers to actually reach some slow speeds representing only several percents from the SH-BAW speed $V_{tem}$. This possibility and the structural geometry when plates are used can be called for further miniaturization of different technical devices, for instance, delay lines based on the new dispersive SH-wave.

It is necessary to state right away that no nondispersive Zakharenko waves (Zakharenko, 2005a; Zakharenko, 2005b; Zakharenko, 2007) were recorded in the present study. This fact differs this study from the other recently carried out by Zakharenko (2015a). The nondispersive Zakharenko waves mathematically represent all the extreme points in dispersion relation curves and can be met in many layered (Zakharenko, 2005a; Zakharenko, 2005b) and quantum (Zakharenko, 2005b; Zakharenko, 2007) systems. In the physical sense, these waves represent the dispersion loss within a dispersive wave mode and divide this dispersive wave mode into one, two, or several sub-modes (modes) of dispersive waves with different dispersion types. Indeed, one can find that Figures 1 and 2 do not demonstrate any extreme points.

Figure 2 shows the complicated dispersion relations (the fundamental modes for the case of $V_{new39} < V_{tem}$ and $V_{new40} < V_{tem}$) related to cases (50) and (55). This figure demonstrates the behaviors of the second pair of the new SH-wave velocities $V_{new39}$ and $V_{new40}$, namely their changes versus the normalized value of the half-thickness $kd$. Figures 2a and 2b show the dispersion relations for PZT-5H–Terfenol-D composite and Figures 2c and 2d show the ones for BaTiO$_3$–CoFe$_2$O$_4$ composite. For both the used composites, the complicated dispersion relations shown in Figures 2a and 2c for different values of the normalized parameter $\alpha^2/\varepsilon\mu$ need a clarification for the reader. Let's refer to Figure 3 that graphically shows the nondispersive SH-SAWs in order to better understand the dispersion relations. Figure 3 shows the normalized velocities $V_{BGM}/V_{tem}$ (dotted lines), $V_{new9}/V_{tem}$ (thin lines) and $V_{new11}/V_{tem}$ (thick lines) in dependence on the normalized parameter $\alpha^2/\varepsilon\mu$ for analysis and comparison. It is clearly seen in Figure 3 that the new SH-SAW velocity $V_{new9}$ studied in (Zakharenko, 2015a) can cross the surface Bleustein-Gulyaev-Melkumyan wave velocity $V_{BGM}$. This occurs at some value of $\alpha^2/\varepsilon\mu = (\alpha^2/\varepsilon\mu)_{BGM}$ (Zakharenko, 2015a). Concerning the other new SH-SAW velocity $V_{new11}$, it can touch the SH-BAW velocity $V_{tem}$ at two values of the parameter $\alpha^2/\varepsilon\mu$. It was found in numerical experiments that the first touch ($V_{new11} = V_{tem}$) occurs at the same value of $\alpha^2/\varepsilon\mu = (\alpha^2/\varepsilon\mu)_{BGM}$. For this study, it is convenient to denote this value by $(\alpha^2/\varepsilon\mu)_{tem1}$ to further operate with it, where $(\alpha^2/\varepsilon\mu)_{tem1} = (\alpha^2/\varepsilon\mu)_{BGM}$. The second touch ($V_{new11} = V_{tem}$) clearly illuminated in figure 3b takes place when $\alpha^2/\varepsilon\mu \to 1$. It is natural to denote it by $(\alpha^2/\varepsilon\mu)_{tem2}$. It is flagrant that at these two values, the new SH-SAW with velocity $V_{new11}$ cannot propagate because there is $V_{new11} = V_{tem}$. This means that here the SH-BAW becomes stable and it is hard to cause any instability in order to get the SH-SAW propagation. So, these two values of $(\alpha^2/\varepsilon\mu)_{tem1}$ and $(\alpha^2/\varepsilon\mu)_{tem2}$ play their important roles in the dispersion relations shown in

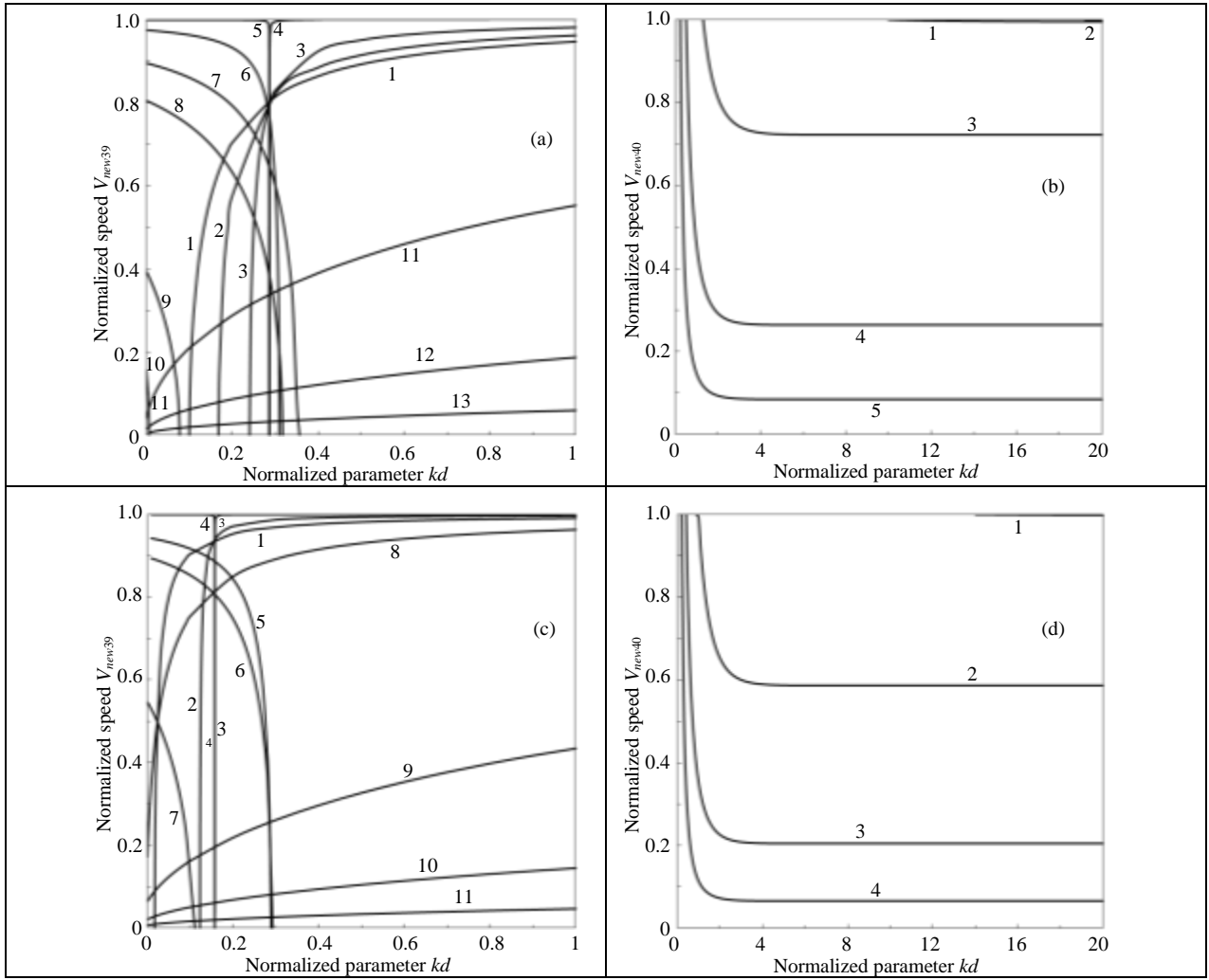

Fig. 2. The normalized velocities $V_{new39}/V_{tem}$ (formula (50)) and $V_{new40}/V_{tem}$ (formula (55)) of the fundamental modes of the dispersive SH-waves propagating in the PEM plates versus the normalized value of the half-thickness $kd$: (a) $V_{new39}$ in PZT-5H–Terfenol-D, where (1) $\alpha^2/\varepsilon\mu = 1.0 \times 10^{-6}$, (2) 0.04, (3) 0.16, (4) 0.2809, (5) 0.281961, (6) 0.36, (7) 0.64, (8) 0.81, (9) 0.9801, (10) 0.998001, (11) 0.99980001, (12) 0.9999800001, (13) 0.999998000001; (b) $V_{new40}$ in PZT-5H–Terfenol-D, where (1) $\alpha^2/\varepsilon\mu = 1.0 \times 10^{-8}$, (2) 0.09, (3) 0.99980001, (4) 0.9999800001, (5) 0.999998000001; (c) $V_{new39}$ in BaTiO$_3$–CoFe$_2$O$_4$, where (1) $\alpha^2/\varepsilon\mu = 1.0 \times 10^{-8}$, (2) 0.09, (3) 0.159201, (4) 0.16, (5) 0.64, (6) 0.81, (7) 0.9801, (8) 0.998001, (9) 0.99980001, (10) 0.9999800001, (11) 0.999998000001; (d) $V_{new40}$ in BaTiO$_3$–CoFe$_2$O$_4$, where (1) $\alpha^2/\varepsilon\mu = 0.998001$, (2) 0.99980001, (3) 0.9999800001, (4) 0.999998000001.

Figures 2a and 2c. Indeed, these two values actually split the dependence $V_{new11}(\alpha^2/\varepsilon\mu)$ into three following parts for either studied composite: the first part is confined between $\alpha^2/\varepsilon\mu > 0$ and $\alpha^2/\varepsilon\mu < (\alpha^2/\varepsilon\mu)_{tem1}$, the second is for the following interval $(\alpha^2/\varepsilon\mu)_{tem1} < \alpha^2/\varepsilon\mu < (\alpha^2/\varepsilon\mu)_{tem2}$, and the last is for $\alpha^2/\varepsilon\mu > (\alpha^2/\varepsilon\mu)_{tem2}$ with the limitation of $\alpha^2/\varepsilon\mu < 1$. It is necessary to state right away that the first part represents a practical interest because the value of $\alpha^2/\varepsilon\mu$ is generally very small. However, it is possible that for some PEM composites, the value of $(\alpha^2/\varepsilon\mu)_{tem1}$ can be also small enough and it is possible to reach the second part for investigations. For the studied composites, it is useful to give these two threshold values of $\alpha^2/\varepsilon\mu$:

$(\alpha^2/\varepsilon\mu)_{tem1} \sim 0.2809$ and $(\alpha^2/\varepsilon\mu)_{tem2} \sim 0.998940281$ for PZT-5H–Terfenol-D, $(\alpha^2/\varepsilon\mu)_{tem1} \sim 0.159201$ and $(\alpha^2/\varepsilon\mu)_{tem2} \sim 0.997581464$ for BaTiO$_3$–CoFe$_2$O$_4$. Let's return to the dispersion relations shown in Figures 2a and 2c. For either studied composite, the dispersion relations for $\alpha^2/\varepsilon\mu < (\alpha^2/\varepsilon\mu)_{tem1}$ represent curves of the $V_{new39}$ versus the parameter $kd$. These dependencies begin with the $V_{new39} = 0$ at some corresponding values of $kd = (kd)_{th} > 0$. This means that there is a "silence" zone for the normalized velocity $V_{new39}/V_{tem}$ for $0 < kd < (kd)_{th}$. Then, $kd \to \infty$ leads to $V_{new39} \to V_{new11}$. Therefore, it is possible to state that the dispersive SH-wave velocity $V_{new39}$ as the function of $kd$ always increases from its zero value up to $V_{new39} = V_{new11}$. This means that this case is for the dispersion type $V_g > V_{ph}$, where $V_{ph}$ and $V_g$ are the phase and group velocities, respectively. In this case, $V_{ph}$ represents $V_{new39}$.

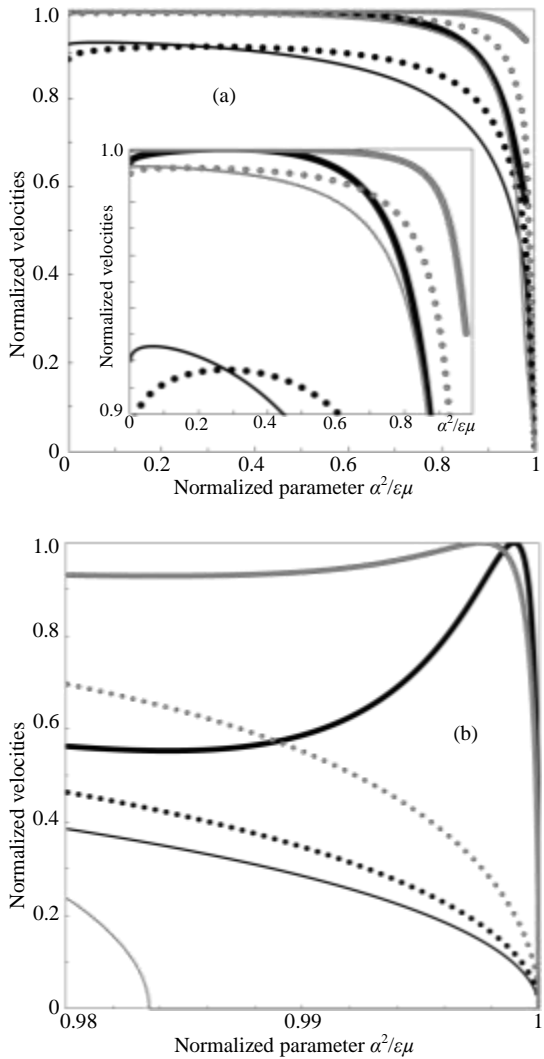

Fig. 3. The normalized velocities $V_{BGM}/V_{tem}$ (dotted lines), $V_{new9}/V_{tem}$ (thin lines) and $V_{new11}/V_{tem}$ (thick lines) of the nondispersive SH-waves propagating in the PEMs such as PZT-5H–Terfenol-D (black lines) and BaTiO$_3$–CoFe$_2$O$_4$ (gray lines) versus the normalized parameter $\alpha^2/\varepsilon\mu$. (a) The main dependencies are shown, where the insertion demonstrates the crossing points between the velocities $V_{BGM}$ and $V_{new9}$. (b) The complicated dependencies of $V_{new9}/V_{tem}$ and $V_{new11}/V_{tem}$ for $\alpha^2/\varepsilon\mu \to 1$ are shown.

The other dispersion type, namely $V_g < V_{ph}$ exists for $(\alpha^2/\varepsilon\mu)_{tem1} < \alpha^2/\varepsilon\mu < (\alpha^2/\varepsilon\mu)_{tem2}$. For these values of $\alpha^2/\varepsilon\mu$, the velocity $V_{new39}$ representing the $V_{ph}$ always decreases from some maximum value at $kd = 0$ to $V_{new39} = 0$ at some finite value of $kd = (kd)_{cut} > 0$. This manifests that there is a cut-off for each fundamental mode corresponding to values of $\alpha^2/\varepsilon\mu$ satisfying the case of $(\alpha^2/\varepsilon\mu)_{tem1} < \alpha^2/\varepsilon\mu < (\alpha^2/\varepsilon\mu)_{tem2}$. As soon as the value of $\alpha^2/\varepsilon\mu$ is passed into the third part with $(\alpha^2/\varepsilon\mu)_{tem2} < \alpha^2/\varepsilon\mu < 1$, the first dispersion type of $V_g > V_{ph}$ is actual anew. For $(\alpha^2/\varepsilon\mu)_{tem2} < \alpha^2/\varepsilon\mu < 1$, the velocity $V_{new39}$ commences with some nonzero value at $kd = 0$ and approaches the velocity $V_{new11}$ at $kd \to \infty$. The fundamental modes for this third case look like there is neither a "silence" zone nor a cut-off. It is now possible to make a statement that at these two threshold values of $(\alpha^2/\varepsilon\mu)_{tem1}$ and $(\alpha^2/\varepsilon\mu)_{tem2}$ with $V_{new11} = V_{tem}$, the dispersion type is changed. This is similar to the nondispersive Zakharenko waves discussed above. However, one deals here with two different mechanisms for the phenomenon of $V_{new11} = V_{tem}$ and the phenomenon called the nondispersive Zakharenko waves. For the first phenomenon there is a break when $V_{new11} = V_{tem}$ leading to the change of the dispersion type. On the other hand, the nondispersive Zakharenko waves represent extreme points inside the same dispersive wave mode and at these extreme points, the dispersion types are changed in a smooth way with no break. This is the main difference between two mechanisms of dispersion type changes.

It is also possible to discuss the dispersion relations shown in Figures 2a and 2c in the vicinity of $(\alpha^2/\varepsilon\mu)_{tem1}$. This is the threshold value of $\alpha^2/\varepsilon\mu$ at which the corresponding fundamental mode looks like a δ-function: for the value of $\alpha^2/\varepsilon\mu$ being slightly below the threshold value $(\alpha^2/\varepsilon\mu)_{tem1}$, the value of the velocity $V_{new39}$ jumps from zero value up to the SH-BAW speed $V_{tem}$ (curves (4) and (3) in Figures 2a and 2c, respectively) and for the case when the value of $\alpha^2/\varepsilon\mu$ is slightly above $(\alpha^2/\varepsilon\mu)_{tem1}$ there is a reverse situation because the value of the velocity $V_{new39} \sim V_{tem}$ falls down to its zero value at the cut-off parameter $(kd)_{cut} > 0$, see curves (5) and (4) in Figures 2a and 2c, respectively. This fact can mean that a very gentle increase (decrease) in, for instance, an external magnetic field can allow us to record in experiments the new dispersive SH-wave propagation and dramatic changes in the propagation velocity from zero to the SH-BAW speed $V_{tem}$. These PEM-SH-wave fundamental modes look like a soliton kink for $\alpha^2/\varepsilon\mu < (\alpha^2/\varepsilon\mu)_{tem1}$ and a soliton anti-kink for $\alpha^2/\varepsilon\mu > (\alpha^2/\varepsilon\mu)_{tem1}$. It is even possible to say that there is a singular dark shape soliton for $\alpha^2/\varepsilon\mu = (\alpha^2/\varepsilon\mu)_{tem1}$. "Soliton" is the reserved word for a sole wave known already during a century in the physics-chemistry-biology of nonlinear processes including the "nanoworld" that is popular for extensive study in the last decades. The evidence of the soliton-like (quasi-particle) behavior of some fundamental modes of purely mechanical waves was reported in theoretical work (Zakharenko, 2005c; Zakharenko, 2010b) regarding to the two-layer structure consisting of a layer on a substrate.

Finally, it is allowed to discuss the last discovered new dispersive PEM-SH-waves propagating with the velocity $V_{new40}$. Figures 2b and 2d show the fundamental modes in the case of $V_{new40} < V_{tem}$. The dispersion curves in these Figures look relatively simpler compared with the dispersive behavior of the velocity $V_{new39}$. However, the existence of the fundamental modes in the studied case of $V_{new40} < V_{tem}$ depends on the aforementioned threshold values $(\alpha^2/\varepsilon\mu)_{tem1}$ and $(\alpha^2/\varepsilon\mu)_{tem2}$. For $\alpha^2/\varepsilon\mu < (\alpha^2/\varepsilon\mu)_{tem1}$ there are weakly dispersive SH-waves localized just below the SH-BAW speed $V_{tem}$, $V_{new40} \sim V_{tem}$. This case is shown by curve (1) in Figure 2b for the PZT-5H–Terfenol-D composite. For the other composite there also exist such weakly dispersive waves for the values of $\alpha^2/\varepsilon\mu < (\alpha^2/\varepsilon\mu)_{tem1}$. However, they were not shown in Figure 2d because the fundamental modes start at large values of $kd > 20$, for instance, $kd \sim 54.1$ for $\alpha^2/\varepsilon\mu = 10^{-10}$ and $kd \sim 54.3$ for $\alpha^2/\varepsilon\mu = 10^{-6}$. These two values of $kd$ are close to each other because here $V_{new40} \sim V_{tem}$ always occurs. In this case of $\alpha^2/\varepsilon\mu < (\alpha^2/\varepsilon\mu)_{tem1}$ one deals with the dispersion type $V_g < V_{ph}$ because the velocity $V_{ph} = V_{new40}$ decreases. It is worth mentioning here that the dispersion type must change at the threshold value $(\alpha^2/\varepsilon\mu)_{tem1}$. For the second case of $(\alpha^2/\varepsilon\mu)_{tem1} < \alpha^2/\varepsilon\mu < (\alpha^2/\varepsilon\mu)_{tem2}$ there is no any dispersion curve with the other dispersion type $V_g > V_{ph}$. This is the peculiarity for the new dispersive SH-waves defined by dispersion relation (55). This dispersion type perhaps exists but not in the studied case of $V_{new40} < V_{tem}$. Thus, it is possible to state that all the dispersion curves shown in Figures 2b and 2d pertain to the third case of $(\alpha^2/\varepsilon\mu)_{tem2} < \alpha^2/\varepsilon\mu < 1$, but curve (1) in Figure 2b. The dispersion type is here $V_g < V_{ph}$ anew because it must be changed at $\alpha^2/\varepsilon\mu = (\alpha^2/\varepsilon\mu)_{tem2}$. So, complicated dispersion relations (50) and (55) shown in Figures 1 and 2 were graphically investigated and found peculiarities were briefly discussed. This allows the author to make the conclusive statement below.

## CONCLUSION

In summary it is possible to point out that this theoretical study has discovered four dispersive new SH-waves. The discovered new waves can propagate in the transversely isotropic (6 *mm*) piezoelectromagnetic plates. The studied case relates to the homogeneous boundary conditions applied for both the upper and lower plate surfaces: $\sigma_{32} = 0$, $\varphi = \varphi^f$, $D = D^f$, $\psi = \psi^f$, and $B = B^f$. The obtained dispersion relations are shown in Figures 1 and 2 for two different composites: PZT-5H–Terfenol-D and BaTiO$_3$–CoFe$_2$O$_4$. These samples were chosen due to the fact that their characteristics are significantly different from each other and they were also previously treated in previous theoretical investigations that allows for comparison. The theoretical study developed in this paper has demonstrated and discussed some complicated peculiarities that can reveal some trying features of the magnetoelectric effect and can be used for constitution of different technical devices. This study can be also useful for the problems of the nondestructive testing and evaluation of PEM plates. It is obvious that plates possessing 2D geometry are apt for further miniaturization of various technical devices: dispersive delay lines, switches, etc. It is expected that slower speeds can be preferable, for instance, in dispersive delay line devices. Indeed, various devices can be called for both conventional electronics and spintronics.

## REFERENCES

Al'shits, VI., Darinskii, AN. and Lothe, J. 1992. On the existence of surface waves in half-infinite anisotropic elastic media with piezoelectric and piezomagnetic properties. Wave Motion. 16(3):265-283.

Auld, BA. 1990. Acoustic Fields and Waves in Solids. Krieger Publishing Company (vol. I and II, 2$^{nd}$ ed.). pp878.

Bleustein, JL. 1968. A new surface wave in piezoelectric materials. Applied Physics Letters. 13(12):412-413.

Dieulesaint, E. and Royer, D. 1980. Elastic waves in solids: Applications to signal processing. J. Wiley, New York, USA. (Translated by Bastin, A. and Motz, M., Chichester). pp511.

Durdag, K. 2009. Wireless surface acoustic wave sensors. Sensors and Transducers Journal. 106(7):1-5.

Fiebig, M. 2005. Revival of the magnetoelectric effect. Journal of Physics D: Applied Physics. 38(8):R123-R152.

Gulyaev, YuV. 1969. Electroacoustic surface waves in solids. Soviet Physics Journal of Experimental and Theoretical Physics Letters. 9(1):37-38.

Gulyaev, YuV. 1998. Review of shear surface acoustic waves in solids. IEEE Transactions on Ultrasonics, Ferroelectrics, and Frequency Control. 45(4):935-938.

Hirao, M. and Ogi, H. 2003. EMATs for science and industry: Non-contacting ultrasonic measurements. Boston, MA, Kluwer Academic Publishers. pp363.

Kimura, T. 2012. Magnetoelectric hexaferrites. Annual Review of Condensed Matter Physics. 3(1):93-110.

Lardat, C., Maerfeld, C. and Tournois, P. 1971. Theory and performance of acoustical dispersive surface wave delay lines. Proceedings of the IEEE. 59(3):355-364.

Melkumyan, A. 2007. Twelve shear surface waves guided by clamped/free boundaries in magneto-electro-elastic materials. International Journal of Solids and Structures. 44(10):3594-3599.

Özgür, Ü., Alivov, Ya. and Morkoç, H. 2009. Microwave ferrites, part 2: Passive components and electrical tuning. Journal of Materials Science. Materials in Electronics. 20(10):911-952.

Pullar, RC. 2012. Hexagonal ferrites: A review of the synthesis, properties and applications of hexaferrite ceramics. Progress in Materials Science. 57(7):1191-1334.

Ribichini, R., Cegla, F., Nagy, PB. and Cawley, P. 2010. Quantitative modeling of the transduction of electromagnetic acoustic transducers operating on ferromagnetic media. IEEE Transactions on Ultrasonics, Ferroelectrics, and Frequency Control. 57(12):2808-2817.

Srinivasan, G. 2010. Magnetoelectric composites. Annual Review of Materials Research. 40(1):153-178.

Thompson, RB. 1990. Physical principles of measurements with EMAT transducers. In: Physical Acoustics. Eds. Mason WP. and Thurston, RN. Academic Press, New York, USA. 19:157-200.

Zakharenko, AA. 2005[a]. Dispersive Rayleigh type waves in layered systems consisting of piezoelectric crystals bismuth silicate and bismuth germanate. Acta Acustica united with Acustica. 91(4):708-715.

Zakharenko, AA. 2005[b]. Different dispersive waves of bulk elementary excitations in bulk superfluid helium–II at low temperatures. In: CD-ROM Proceedings of the Forum Acusticum. Budapest, Hungary. L79-L89.

Zakharenko, AA. 2005[c]. Analytical studying the group velocity of three-partial Love (type) waves in both isotropic and anisotropic media. Nondestructive Testing and Evaluation. 20(4):237-254.

Zakharenko, AA. 2007. Different Zakharenko waves in layered and quantum systems. In: Conference Proceedings of the International Commission for Acoustics. ICA2007 Congress, Madrid, Spain. pp4. http://www.sea-acustica.es/WEB_ICA_07/fchrs/papers/phy-08-018.pdf

Zakharenko, AA. 2010[a]. Propagation of seven new SH-SAWs in piezoelectromagnetics of class 6 *mm*. LAP LAMBERT Academic Publishing GmbH & Co. KG, Saarbruecken-Krasnoyarsk. pp84.

Zakharenko, AA. 2010[b]. Slow acoustic waves with the anti-plane polarization in layered systems. International Journal of Modern Physics B. 24(4):515-536.

Zakharenko, AA. 2011. Analytical investigation of surface wave characteristics of piezoelectromagnetics of class 6 *mm*. ISRN Applied Mathematics (India). 2011:408529. pp8.

Zakharenko, AA. 2012[a]. Thirty two new SH-waves propagating in PEM plates of class 6 *mm*. LAP LAMBERT Academic Publishing GmbH & Co. KG, Saarbruecken-Krasnoyarsk. pp162.

Zakharenko, AA. 2012[b]. On wave characteristics of piezoelectromagnetics. Pramana – Journal of Physics. 79(2):275-285.

Zakharenko, AA. 2013[a]. New nondispersive SH-SAWs guided by the surface of piezoelectromagnetics. Canadian Journal of Pure and Applied Sciences. 7(3):2557-2570.

Zakharenko, AA. 2013[b]. Piezoelectromagnetic SH-SAWs: A review. Canadian Journal of Pure & Applied Sciences. 7(1):2227-2240.

Zakharenko, AA. 2013[c]. Peculiarities study of acoustic waves' propagation in piezoelectromagnetic (composite) materials. Canadian Journal of Pure and Applied Sciences. 7(2):2459-2461.

Zakharenko, AA. 2013[d]. Fundamental modes of new dispersive SH-waves in piezoelectromagnetic plate. Pramana – Journal of Physics. 81(5):819-827.

Zakharenko, AA. 2013[e]. Consideration of SH-wave fundamental modes in piezoelectromagnetic plate: Electrically open and magnetically open boundary conditions. Waves in Random and Complex Media. 23(4):373-382.

Zakharenko, AA. 2014[a]. Some problems of finding of eigenvalues and eigenvectors for SH-wave propagation in transversely isotropic piezoelectromagnetics. Canadian Journal of Pure and Applied Sciences. 8(1):2783-2787.

Zakharenko, AA. 2014[b]. Investigation of SH-wave fundamental modes in piezoelectromagnetic plate: Electrically closed and magnetically closed boundary conditions. Open Journal of Acoustics. 4(2):90-97.

Zakharenko, AA. 2015[a]. On new dispersive SH-waves propagating in piezoelectromagnetic plates. Open Journal of Acoustics. 5(3):122-137.

Zakharenko, AA. 2015[b]. Dramatic influence of the magnetoelectric effect on the existence of the new SH-SAWs propagating in magnetoelectroelastic composites. Open Journal of Acoustics. 5(3):73-87.

Zakharenko, AA. 2015[c]. A study of new nondispersive SH-SAWs in magnetoelectroelastic medium of symmetry class 6 *mm*. Open Journal of Acoustics. 5(3):95-111.